\documentclass[%
 reprint,
 prb,
 amsmath,amssymb,
 aps,
 longbibliography
]{revtex4-1}
\usepackage{amsmath,amssymb}
\usepackage{bm}
\usepackage[dvipdfmx]{graphicx}
\usepackage{float}
\usepackage{amsthm}
\usepackage{mathtools}
\usepackage[driverfallback=dvipdfmx,colorlinks=true,
linkcolor=blue,citecolor=blue]{hyperref}
\newcommand{\average}[1]{\ensuremath{\langle#1\rangle} }
\begin{document}
\title{Diffusive real-time dynamics of a particle with Berry curvatures}

\author{Kou Misaki,${}^{1}$ Seiji Miyashita,${}^{2}$ Naoto Nagaosa${}^{1,3}$}

\affiliation{
$^1$Department of Applied Physics, The University of Tokyo, 
Bunkyo, Tokyo 113-8656, Japan
\\
$^2$Department of Physics, The University of Tokyo, Bunkyo, Tokyo 113-8656, Japan
\\
$^3$RIKEN Center for Emergent Matter Science (CEMS), Wako, Saitama 351-0198, Japan
\\
}
\date{\today}
\begin{abstract}
We study theoretically the influence of Berry phase on the real-time
dynamics of the single particle focusing on the diffusive dynamics, i.e.,
the time-dependence of the distribution function. Our model can be
applied to
the real-time dynamics of intraband relaxation and diffusion of optically
excited excitons, trions or particle-hole pair.
We found that the dynamics at the early stage is deeply
influenced by the Berry curvatures in real-space ($B$), momentum-space
($\Omega$),
and also the crossed space between these two ($C$).
For example, it is found that $\Omega$ induces the rotation of the
wave packet and causes the
time-dependence of the mean square displacement of the particle
to be linear in time $t$ at the initial stage; it is qualitatively
different from the $t^3$ dependence in the absence of the Berry curvatures.
It is also found that $\Omega$ and $C$ modifies the characteristic
time scale of the thermal equilibration of momentum distribution.
Moreover, the dynamics
under various combinations of $B$, $\Omega$ and $C$
shows singular behaviors such as the critical slowing
down or speeding up of the momentum  equilibration and the reversals of
the direction of rotations. The relevance of our model
for time-resolved experiments
in transition metal dichalcogenides is also discussed.
\end{abstract}
\maketitle

\section{Introduction}
The role of Berry phase \cite{berry1984quantal}
in wave mechanics has been attracting
intensive attention. The effects from both the geometry
characterized by the Berry curvature, which can be understood as a
modification of commutation relations between phase space coordinates
\cite{xiao2005berry,duval2006berry}, and
its global aspects captured by the topological indices are the focus
of recent studies. The former includes the anomalous Hall effect
\cite{nagaosa2010anomalous},
spin Hall effect \cite{murakami2003dissipationless,sinova2004universal},
and magnon Hall effect \cite{onose2010observation}, while the topological
insulators and topological superconductors are the examples of the
latter \cite{hasan2010colloquium,qi2011topological}.
Berry phase has been discussed for
the ground states and the linear responses near the thermal
equilibrium
\cite{nagaosa2010anomalous,murakami2003dissipationless,sinova2004universal,onose2010observation}, 
the general cyclic evolution of a quantum state
\cite{aharonov1987phase}, and the periodically driven systems
\cite{wilczek1989geometric,young2012first1,young2012first2,morimoto2016topological}.

On the other hand, the role of Berry phase in the
real-time dynamics far from the equilibrium has
been less studied.  Especially the diffusion processes
\cite{olson2007nonequilibrium} are
fundamental for propagation of particles, chemical reactions, and even
biological phenomena \cite{van1992stochastic}.
Especially, the real-time dynamics becomes a
tractable issue experimentally
due to the technological developments, e.g.,
ultra-fast time-resolved spectroscopies in cold atom systems
\cite{dalibard2011colloquium,goldman2014light} and in
solids \cite{krausz2009attosecond,goulielmakis2010real}.
Although there have been some proposals and
experiments in cold atom systems \cite{alba2011seeing,zhao2011chern,price2012mapping,aidelsburger2015measuring,
duca2015aharonov,flaschner2016experimental,price2016measurement} and
photonic lattice systems \cite{ozawa2014anomalous}
for measuring the Berry curvatures in momentum space,
the diffusive dynamics has not been explored.

In this work, we study the role of Berry phase in diffusion processes
\cite{ao2004potential,van1992stochastic}.
We consider the Berry curvatures in real-space ($B$), 
momentum-space ($\Omega$), 
and also the crossed space between these two ($C$). 
These three curvatures play distinct roles in the real-time dynamics 
of diffusion starting from the initial condition of fixed 
position and momentum. Therefore, the results offer yet
another method 
to disentangle the Berry curvatures in terms of time-resolved 
experiments. Also it is found that the interference between
them results in rich phenomena including 
the singular behaviors as shown below.

\section{Model and results}
\subsection{Semiclassical stochastic equation}
The semiclassical equation for the wave packet localized both in
position and momentum space is, if we include
the friction and fluctuation caused by a heat bath
\cite{ao2004potential,yin2006existence,olson2007nonequilibrium} (see
Appendix C for derivation),
\begin{align}
\dot{r}_{i}&=\frac{\partial \epsilon(\bm{r},\bm{p})}{\partial p_{i}}-\left(
 (\hat{\Omega}_{pp})_{ij}\dot{p}_{j}+(\hat{\Omega}_{pr})_{ij}\dot{r}_{j}\right),\label{Langer}\\
\dot{p}_{i}&=-\frac{\partial \epsilon(\bm{r},\bm{p})}{\partial r_{i}}+
\left(
  (\hat{\Omega}_{rp})_{ij}\dot{p}_{j}+(\hat{\Omega}_{rr})_{ij}\dot{r}_{j}\right)\nonumber\\
  &\quad  -m\gamma
  \dot{r}_{i}+\sqrt{2m\gamma k_BT}\xi_i(t),\label{Langep}
\end{align}
where $m$ is the mass of the particle, $\gamma$ is the friction
constant, $k_B$ is the Boltzmann constant, $T$ is the temperature of
the system, $i,j=1,\dots, d$ and $d$ is the spatial dimension of the
system. $(\hat{\Omega}_{XX})_{\alpha\beta}$ ($\bm{X}=(\bm{r},\bm{p})$, and
$\alpha,\beta=1,\dots, 2d$ are the coordinates of phase space.) is
the Berry curvature, and $\epsilon(\bm{r},\bm{p})$ is the
energy of the particle.
$\xi_i(t)$ is the
Gaussian fluctuation force and satisfies
$\average{\xi_i(t)\xi_j(t')}=\delta(t-t')\delta_{ij}$ and
$\average{\xi_i(t)}=0$,
where the bracket denotes the ensemble average.

From now on, we will assume that the spatial dimension $d=2$,
$\epsilon(\bm{p})=\bm{p}^2/(2m)$, $\hat{\Omega}_{pp}=(\Omega/\hbar) i\sigma_y$,
$\hat{\Omega}_{rr}=qB i\sigma_y$ and $\hat{\Omega}_{rp}=C I_2$,
where $q$ is the charge of the particle and $I_2$ and $\sigma_y$ are
$2\times 2$ unit matrix and $y$ component of Pauli matrices,
respectively. We set $\hbar=1$ henceforth.
Here we assumed $\Omega$, $qB$ and $C$ to be constant.
We defer the discussion for the applicability of our model to
real experiments to the end of the paper.
Here,  $B$ and $\Omega$ are the real space magnetic field
perpendicular to our two dimensional system and Berry curvature
in momentum space, respectively. As for $C$, in the presence of
elastic deformation field $u_i(\bm{r})$, $\hat{\Omega}_{rp}$ can be
calculated as \cite{sundaram1999wave}
\begin{equation}
 (\hat{\Omega}_{rp})_{ij}=\frac{\partial u_j}{\partial r_i}
  \left(1-\frac{m_0}{m} \right)\eqqcolon w_{ji}
  \left(1-\frac{m_0}{m} \right),
\end{equation}
where $m_0$ is an unrenormalized bare mass of the particle.
Here we restrict our attention to the symmetric part of
$w_{ij}$ ($w^s_{ij}=\frac{1}{2} ( w_{ij}+ w_{ji})$). If we
consider the case where the system is under the uniform, isotropic
and weak pressure, according to
Hooke's law \cite{landau1986theory},
$w^s_{ij}\propto \delta_{ij} $. Moreover, for the system with $m\ll
m_0$, small amount of deformation leads to large $C \sim 1$.

As we mentioned in the introduction, varying Berry curvatures
amount to modifying the commutation relation.
We will see, at particular parameter range, i.e.,
$C=1$ and $qB\Omega=1$, the dynamics becomes
singular. It can be attributed to the singularity of the commutation
relation of the dynamics.
For example, $C$=1 indicates that the $r$ and $p$ commute each other and both
can be determined simultaneously, i.e., the uncertain principle does
not apply in this case.

\begin{figure}
  \centering
   \includegraphics[width=\hsize]{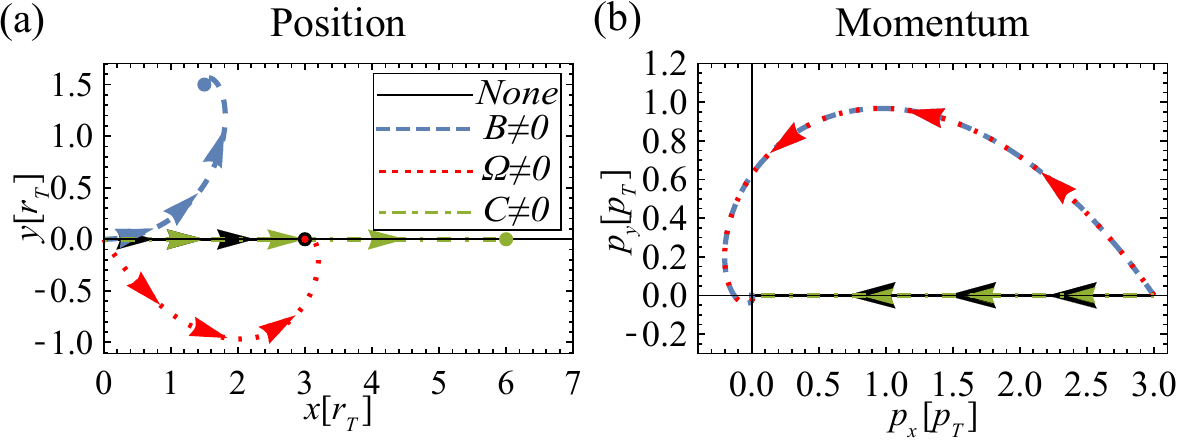}
 \caption{
   The plots for the time evolutions of the averages of the position
 (a) and momentum (b), measured in units of
 $p_T=\sqrt{2m k_BT}$ and $r_T=p_T/(m\gamma)$.
   The initial condition is
   $P(\bm{X},t=0)=\prod_{\alpha}\delta(X_{\alpha}-X^0_{\alpha})$ and
   $(r_{x0},r_{y0},p_{x0},p_{y0})=(0,0,3p_T,0)$.
   The ``None'', ``$B\neq 0$'', ``$\Omega \neq 0$'' and ``$C\neq 0$'' lines are
 the behaviors at dimensionless parameters
 $(qB/(m\gamma),m\gamma\Omega/\hbar,C)=(0,0,0),$ $(-1,0,0),$
 $(0,1,0)$ and $(0,0,-1)$,
 respectively, as is shown in the inset of (a).
 The final position of the particle is denoted by the dots in (a).
 Note that the endpoints of ``None'' and ``$\Omega \neq 0$'' line in (a),
 $B\neq 0$ and $\Omega\neq 0$ line in (b), and ``None'' and
 $C\neq 0$ line in (b), coincide.
 The momentum relaxes to $0$ by friction for all the cases.
 In the case of
 $\Omega\neq 0$, the directions of $\average{p_i}$ and
 $\frac{d}{dt}\average{r_i}$ do not coincide; to see this in the figure,
 we note that, although the initial momentum is purely $x$ direction, the
 initial $\dot{\vec{r}}$ contains $y$ component, because of the finite
 anomalous velocity.}
 \label{COM}
 \end{figure}

\subsection{Fokker-Planck equation and real-time dynamics of diffusion}
From the Langevin equations (\ref{Langer}) and (\ref{Langep}),
we can derive the Fokker-Planck equation,
which describes the time evolution of the probability distribution
function $P(\bm{X},t)$ (Details of the derivation are in 
 Refs. \onlinecite{ao2004potential,yin2006existence,olson2007nonequilibrium})
 and Appendix B:
\begin{align}
  \frac{\partial P(\bm{X},t)}{\partial
 t}&=(\hat{G})_{\alpha\beta}\nabla_{\alpha}[(\nabla_{\beta}\epsilon)P]\nonumber\\
 &\qquad
 +\frac{k_BT}{2}(\hat{G}+\hat{G}^T)_{\alpha\beta}\nabla_{\alpha}\nabla_{\beta}P,
 \label{FP2}
\end{align}
where the matrix $\hat{G}$ is the inverse of
\begin{equation}
 \hat{G}^{-1}=\left(
\begin{array}{cc}
 m\gamma I_2- qB i\sigma_y&(1-C)I_2 \\
 -(1+C)I_2&-\Omega i\sigma_y \\
\end{array}
	      \right).
\end{equation}
 Here we assumed that the matrix $\hat{G}^{-1}$ is regular:
\begin{equation}
 \det \hat{G}^{-1}=\left[(1-C)^2-qB\Omega\right]^2+(m\gamma\Omega)^2 \neq 0.
  \label{detGinverse}
\end{equation}
 We will discuss what happens if $G^{-1}$ is singular later.

Now we study the time-evolution of the distribution function
$P(\bm{X},t)$. 
Because of the assumption of quadratic dispersion
of $\epsilon(\bm{p})$ and constant Berry
curvatures, we can exactly solve Eq. (\ref{FP2})
with the initial condition of fixed position and momentum
\cite{van1992stochastic}:
$P(\bm{X},t=0)=\prod_{\alpha}\delta(X_{\alpha}-X^0_{\alpha})$,
where $\bm{X}_0=(\bm{r}_0,\bm{p}_0)$ denotes the initial coordinate and
momentum.
Since the solution is the Gaussian distribution, it is enough to
calculate the first and second moments for specifying the
probability distribution.

The time evolution of the first moment is
shown in Fig. \ref{COM} in
the case of only one of $B$, $\Omega$ and $C$ is nonzero.
We define two time scales which characterize the dynamics,
$1/\gamma_1$ and $1/\gamma_2$:
\begin{align}
 \gamma_1&=\frac{(1-C)^2
 \gamma}{[(1-C)^2-qB\Omega]^2+m^2\gamma^2\Omega^2},\label{gamma1}\\
 \gamma_2&=\frac{-qB(1-C)^2+(q^2B^2+m^2\gamma^2)\Omega}
 {m\{[(1-C)^2-qB\Omega]^2+m^2\gamma^2\Omega^2\}},\label{gamma2}
\end{align}
where $1/\gamma_1$ is the
relaxation time toward the final position and momentum,
and $\gamma_2$ represents the
frequency of the characteristic rotational motion.
We can see the
characteristic rotational motion
when $\gamma_2 \neq 0$, i.e., $B \neq 0$ or $\Omega \neq 0$ in
Fig. \ref{COM}.

As for the second moment, we define the correlation function
$\average{\average{X_{\alpha}(t)X_{\beta}(t)}}=
  \average{(X_{\alpha}(t)-\average{X_{\alpha}(t)})
  (X_{\beta}(t)-\average{X_{\beta}(t)})}$. Then
the long time behavior of
$\average{\average{r_ir_j}}$ is, as $t\to \infty$,
\begin{align}
 &\average{\average{r_{i}(t)r_j(t)}}
 =\biggl(\frac{2m\gamma k_BT}{q^2B^2+m^2\gamma^2}t\nonumber\\
 &\quad+\frac{m k_B
 T (1-C)^2}{(q^2B^2+m^2\gamma^2)^2}(q^2B^2-3m^2\gamma^2)+
 \mathcal{O}(e^{-\gamma_1 t})\biggr)\delta_{ij}.
\label{diffconst}
\end{align}

On the other hand, the short time behavior of
$\average{\average{r_ir_i}}$ (no summation) is, as $t\to 0$,
\begin{align}
 &\average{\average{r_{i}(t)r_i(t)}}=
 R_1 t-R_2 t^2+R_3 t^3+\mathcal{O}(t^4),\label{rrshort}
\end{align}
where
\begin{align}
 R_1&=\frac{2m\gamma\Omega^2 k_BT
 }{\text{det}\hat{G}^{-1}},\\
 R_2&=\frac{2(1-C)^2m\gamma^2 \Omega^2 k_BT}
 {(\text{det}\hat{G}^{-1})^2},\\
 R_3&=
 2(1-C)^2\gamma k_BT\nonumber\\
 &\,\times\frac{[(1-C)^2-qB\Omega]^3
 +m^2\gamma^2\Omega^2[3(1-C)^2-qB\Omega]}
 {3m(\text{det}\hat{G}^{-1})^3}.
\end{align}
The correlations of the momenta are,
\begin{equation}
 \average{\average{p_i(t)p_j(t)}}
 =mk_BT\left(1-e^{-2\gamma_1 t}\right)
 \delta_{ij}. \label{ppcor}
\end{equation}
This quantity eventually relaxes to $mk_B T$
with the relaxation time $1/(2\gamma_1)$, since the probability
distribution relaxes to the thermal equilibrium, see Appendix B.

Finally, the
cross-correlations between the position and momentum are,
\begin{equation}
 \average{\average{r_i(t)p_j(t)}}
 =\frac{m k_BT}{q^2B^2+m^2\gamma^2}(f_1(t)
 \delta_{ij}-f_2(t)(i\hat{\sigma}_y)_{ij}),\label{rpcor}
\end{equation}
where $f_1(t)=m\gamma (1-C)+e^{-2\gamma_1 t}m\gamma (1-C)[1
 -2e^{\gamma_1 t}\cos(\gamma_2 t)]$ and
 $f_2(t)=qB (1-C)-e^{-2\gamma_1 t}
 [qB(1-C)-2m\gamma (1-C) e^{\gamma_1 t}\sin(\gamma_2 t)]$.
The antisymmetric correlation of $r_i$ and $p_j$, i.e., the second term in
the right hand side of Eq. (\ref{rpcor}),
represents the orbital angular momentum.
    \begin{figure}
     \centering
    \begin{tabular}{c}
     \begin{minipage}{0.98\hsize}
    \centering
    \includegraphics[width=\hsize]{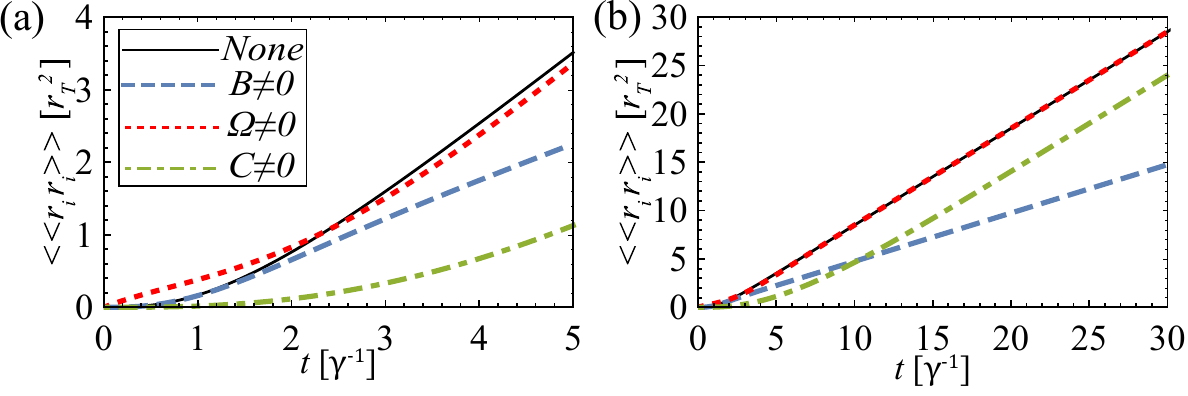}
     \end{minipage}\\
     \begin{minipage}{0.98\hsize}
      \centering
   \includegraphics[width=\hsize]{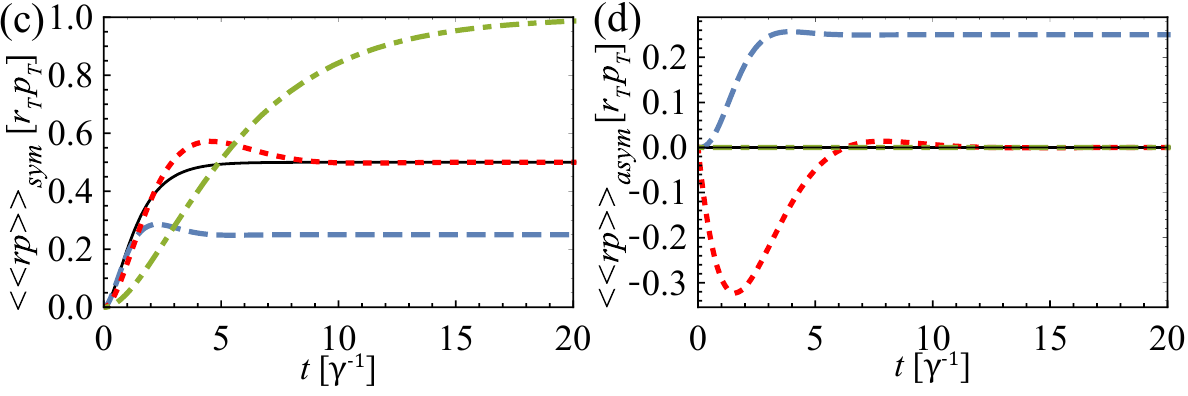}
     \end{minipage}\\
     \begin{minipage}{0.98\hsize}
      \centering
   \includegraphics[width=0.5\hsize]{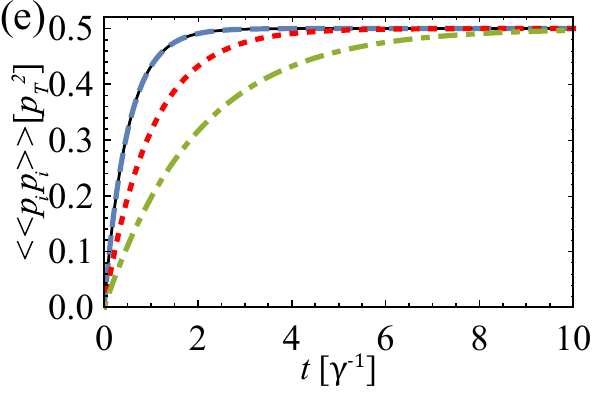}
     \end{minipage}
    \end{tabular}
       \caption{
   The plots for the time evolutions of,
   (a,b)
   $\average{\average{r_ir_i}}$,
    (c) $\average{\average{rp}}_{sym}=
   \average{\average{r_ip_i}}$,
       (d) $\average{\average{rp}}_{asym}=
   \frac{1}{2}(\average{\average{r_xp_y}}-\average{\average{p_xr_y}})$ and
   (e) $\average{\average{p_ip_i}}$ (no summation of
   repeated indices here), where $p_T=\sqrt{2m k_BT}$ and $r_T=p_T/(m\gamma)$.
   Note the differences of the time scale of each panel.
   (a) is the zoom up of (b). The initial condition is
   $P(\bm{X},t=0)=\prod_{\alpha}\delta(X_{\alpha}-X^0_{\alpha})$,
   so $\average{\average{X_{\alpha}X_{\beta}}}=0$ for
   $\forall\alpha,\beta$ at $t=0$.
   The ``None'', ``$B\neq 0$'', ``$\Omega \neq 0$'' and ``$C\neq 0$'' lines are
 the behaviors at dimensionless parameters
 $(qB/(m\gamma),m\gamma\Omega/\hbar,C)=(0,0,0),$ $(-1,0,0),$ $(0,1,0)$ and
 $(0,0,-1)$,
   respectively, as is shown in the inset of (a).
   (a) From Eq. (\ref{rrshort}),
   the short time behavior of \average{\average{r_ir_i}}
   is $\mathcal{O}(t)$ for $\Omega\neq 0$ and $\mathcal{O}(t^3)$ for
   $\Omega=0$.
   (b) We can see the difference of slope and value of
   $\average{\average{r_ir_i}}$ at long time, see Eq. (\ref{diffconst}).
   (c,d) The fact that
   $\frac{1}{2}(\average{\average{r_xp_y}}-\average{\average{p_xr_y}})$
   does not vanish
   indicates the finite angular momentum at long time for
   $B\neq 0$.
  (e) The characteristic relaxation time of
   $\average{\average{p_ip_i}}$ is different in three cases, see
   Eq. (\ref{ppcor}). Note that the ``None'' and ``$B\neq 0$'' line
   coincide in (e).
   These results do not depend on the initial values of
   $X_{\alpha}$, i.e., $X^0_{\alpha}$.}
     \label{correlation}
    \end{figure}

Among the Berry curvatures $B$, $\Omega$ and $C$, the long
time behavior of the diffusive dynamics, i.e., $t >> 1/\gamma_1$,
is characterized mainly
by $B$ and $C$: If $B\neq 0$, the rotational motion from the Lorentz force
(Fig. \ref{correlation}(d))
leads to the slow diffusion, i.e., the small diffusion coefficient,
at long time (Fig. \ref{correlation}(b))
\cite{kurcsunoglu1962brownian,czopnik2001brownian,schutte2014inertia};
the value of
$\average{\average{r_ir_i}}$ (no summation) is affected when $C\neq 0$,
see Eq. (\ref{diffconst}) and Fig. \ref{correlation}(b).
At long time, we do not see any effect of $\Omega$, see Fig.
\ref{correlation}. The reason is that,
after the relaxation of momentum distribution ($t > 1/\gamma_1$), the
force on the particle is balanced and $\dot{p}_i=0$, so the anomalous
velocity term vanishes at the equilibrium of the momentum distribution.
However, the effect of $\Omega$ does appear in
the short time dynamics at $t<1/\gamma_1$.

\subsection{Effect of each Berry curvature on the short time dynamics}
Now we study the effect of individual Berry curvature $B$, $\Omega$, and $C$
on the short time dynamics
by putting only one of them nonzero. The interference between them
will be discussed later.

\noindent --- {\it Real-space magnetic field $B$}

\noindent The rotational motion caused by the Lorentz force
affects the diffusive dynamics. By the rotational motion
(Fig. \ref{correlation}(d)), the diffusion
is suppressed, although $R_3$ in Eq. (\ref{rrshort}) is not affected
by $B$ (Fig. \ref{correlation}(a)). The relaxation of
the momentum distribution is not affected by $B$
(Fig. \ref{correlation}(e)), from Eqs.
(\ref{gamma1}) and (\ref{ppcor}).

\noindent --- {\it Momentum space Berry curvature $\Omega$}

\noindent The anomalous velocity term, combined with
the friction and fluctuation terms in Eq. (\ref{Langep}), result in the
modification of the diffusive dynamics at short time.
Namely, the spread in real space $\average{\average{r_i r_i}}$ becomes fast;
it is linear in $t$ in stark contrast to the usual
$t^3$ behavior without $\Omega$, see Eqs. (\ref{rrshort}),
the definitions of $R_1, R_2,
R_3$ and Fig. \ref{correlation}(a). We note that our model is not a
Smoluchowski equation, which describes the long time scale dynamics and gives $t$
linear behavior in the absence of Berry curvatures.
The coefficient is
$R_1=(2k_BT x^2)/[m\gamma(1+x^2)]$, where $x=m\gamma\Omega$.
The finite angular momentum at short
time can be seen in Fig. \ref{correlation}(d); as we noted above,
this behavior is independent of the initial momentum $p_0$, and can
be understood as the internal rotational motion of the wave packet
of the probability distribution in real space. 
From Eq. (\ref{gamma1}), the characteristic relaxation
time $1/\gamma$ is modified as $1/\gamma_1=
(1 + x^2)/\gamma$, and the relaxation of the momentum
distribution toward the equilibrium becomes slower, see Fig.
\ref{correlation}(e).

\noindent --- {\it Berry curvature in crossed space $C$}

\noindent
The dynamics does not contain the rotational motion, since
$\hat{\Omega}_{rp}$ is the diagonal matrix and
the system is symmetric in the left-handed and right-handed direction.
The effect of $C$ appears in
the modification of the relaxation time
$1/\gamma_1 = (1-C)^2/\gamma$ from Eq. (\ref{gamma1})
(Fig. \ref{correlation}(c,f)) and the diffusion at short time
(Fig. \ref{correlation}(a)) and at long time (Fig. \ref{correlation}(b)).
In particular, for $0<C<1$ ($C<0$), $1/\gamma_1$ is reduced (enhanced)
and the relaxation become faster (slower). When $C=1$, from Eq.
(\ref{detGinverse}) the matrix $\hat{G}^{-1}$ is singular and $\gamma_1$
diverges. We will discuss this singular case below.

\subsection{Interference between Berry curvatures}
Now we consider the effects due to the coexistence of 
different Berry curvatures. In particular, it often happens that
both $qB$ ($C$) and $\Omega$ are finite \cite{xiao2005berry}, e.g.,
when the external magnetic field (the elastic deformation)
is applied to the system with the band structure of finite $\Omega$,
so we discuss these cases.

\noindent --- {\it The interference between $\Omega$ and $B$}

\noindent
Because of the term
$1-qB\Omega$ in the denominator, the presence of both $B$ and $\Omega$
leads to the enhancement of $\gamma_1=\gamma/[(1-qB\Omega)^2+(m\gamma\Omega)^2]$,
which is the reciprocal of the characteristic time scales
of the relaxation. This is in sharp contrast to the case where
only $\Omega$ is finite and the effect is only the reduction of
$\gamma_1$. In particular, if we
regard $\gamma_1$ and $\gamma_2$ as functions of $qB$,
$\gamma_1$ obeys the Lorentzian distribution with a peak
of height $1/(m^2\gamma\Omega^2)$ at
$qB=1/\Omega$ with a half width at half maximum $m\gamma$.
When $qB \Omega=1$ and $\gamma=0$, it is known that the degrees of
freedom of the system is reduced, and
we get the constrained system \cite{duval2000exotic,Faddeev1988}. 
Here, $\gamma$ and $\xi_i(t)$
remove the singularity of the $\det \hat{G}^{-1}$ in Eq.
(\ref{detGinverse}), as was pointed out in Ref.
\onlinecite{olson2007nonequilibrium}.
However, the anomalous behavior appears in the diffusive dynamics:
The minimum of $\gamma_2$ at $B=1/(q\Omega)-m\gamma/(qB)$
($\Omega>0$) dips below zero for $\gamma<1/(2m\Omega)$, and the
characteristic rotational motion for short time changes the sign of the angular
momentum twice as we sweep $B$ from $-\infty$ to $+\infty$.
Since the ratio of peak values of $\gamma_1$ and
$\gamma_2$ is
$|\gamma_{2,\text{peak}}/\gamma_{1,\text{peak}}|=|m\gamma\Omega-1/2|$, if
$m\gamma\Omega\ll 1$, it is possible to detect the rotational motion
before the average of the momentum and position relaxes to the equilibrium.

\noindent --- {\it The interference between $\Omega$ and $C$}

\noindent
In the presence of both $\Omega$ and $C$,
$\gamma_1=[(1-C)^2\gamma]/[(1-C)^2+(m\gamma\Omega)^2]$ and
$\gamma_2=[\gamma(m\gamma\Omega)]/[(1-C)^2+(m\gamma\Omega)^2]$.
From these two quantities, we can see
the resonant behavior as we vary $1-C$, and
this behavior crucially depends on whether $\Omega =0$ or not, as shown below.

When $C=1$ and $\Omega=0$, the dynamics of $p_i$ and $r_i$ completely
decouples, and we get the constraint $p_i=0$. In this case, the system
is governed by the dynamics of $r_i$ only, and we get the Langevin
equation for the Brownian particle. In fact, as $C\to 1$,
$\gamma_1=\gamma/(1-C)^2\to \infty$ and the system
becomes overdamped for all the time
scale.

When $C=1$ and $\Omega\neq 0$,
the singularity of $\hat{G}^{-1}$
is removed, see Eq. (\ref{detGinverse}).
However, the dynamics of $p_i$ and $r_i$ is still decoupled.
As $C\to 1$, we get
$\gamma_1=[(1-C)^2\gamma]/[(1-C)^4+(m\gamma\Omega)^2] \to 0$, and
the system becomes
underdamped for all the time scale, and the effect of the friction and
fluctuation on $p_i$ vanishes.
In this case, we get the singular rotational motion:
The solution of Eq. (\ref{Langep}) is
$(p_x,p_y)=p_0(\cos[t/(m\Omega)+\phi],\sin[t/(m\Omega)+\phi])$
($(p_{x0},p_{y0})=p_0(\cos\phi,\sin\phi)$),
so the dynamics of $p_i$ is purely rotational motion with the frequency
$1/(m\Omega)$ $(=\gamma_2)$, which is singular at $\Omega=0$.
The dynamics of $r_i$ is the same as $\Omega=0$ case discussed above.
Here
we see modification of the commutation relation by the Berry curvatures
decouples the dynamics of $p_i$ and $r_i$.

\section{Discussion}
The results given above offer enough information to determine
Berry curvatures from the measurements of real-time diffusive dynamics.
The relaxation of the momentum distribution is affected in the presence of
``magnetic field'' in momentum space just like
the diffusion coefficient
is modified in the presence of magnetic field in real space, and the
behavior we saw is expected to occur universally also in more complex
models.

As for the coexistence of both $\Omega$ and $B$,
a promising candidate is the surface state of magnetic topological insulator
\cite{onose2010observation,hasan2010colloquium}. 
The exchange gap induced at the surface state leads to the
Berry curvature $\Omega$ and quantized anomalous Hall effect \cite{chang2013experimental}. 
Recently,  it is found that the skyrmions are produced during the
magnetization process of this system \cite{yasuda2016geometric}, which produces the real-space
Berry curvature $B$ due to the scalar spin chirality \cite{nagaosa2013topological}.
In this situation, by tuning the exchange gap and the size of the
skyrmion, the product $q B \Omega$ can be of the order of unity.
Note that the real-space 
Berry curvature produced by the Skyrmion crystal are modulated
spatially, but its effect on the electrons with small wavenumber is
identical to that of the uniform $B$ \cite{Hamamoto2015quantized}.

Even more direct relevance to our model is
the dynamics of optically excited excitons and trions
at $K$ and $K'$ point in transition metal dichalcogenides
\cite{DiXiao}. In this material, when the circularly polarized light is
injected, one can selectively create the bound exciton
at only $K$ or $K'$ point depending on the polarization.
The exchange coupling leads to strong mixing between $K$ and $K'$
excitons, and
the Hamiltonian for the center of mass momentum of excitons
$\vec{k}=k(\cos\phi,\sin\phi)$ is
$H_{\text{D}}=vk(\cos(2\phi)\sigma_x+\sin(2\phi)\sigma_y),$
where $\sigma_i$ is $K$ and $K'$ valley pseudo-spin and
$v \sim 0.79 \,{\rm eV} {\rm \AA}$ represents the mixing from the exchange
coupling \cite{yu2014dirac}.
If we apply magnetic field $B$, by valley
Zeeman effect
\cite{macneill2015breaking,srivastava2015valley,aivazian2015magnetic,li2014valley,wang2015magneto,mitioglu2015optical,stier2016exciton},
the gap $H_{\text{gap}}=\Delta \sigma_z$, where $\Delta \sim 2.3 \,{\rm
meV}$ with $B \sim 10\,{\rm T}$, is induced between $K$ and $K'$
excitons. And if
the temperature is low enough to satisfy $k_T\coloneqq \sqrt{2k_BT\Delta}/v\leq
\Delta/v$, i.e., $ T\leq 13 \,{\rm K}$, Berry curvature can be regarded
as constant $\Omega \sim 1.2\times 10^{5} \,{\rm \AA^2}$
and at the same time the dispersion of the upper band can be approximated as
quadratic.
Also, the authors of Ref. \onlinecite{yu2014dirac}
suggested that binding another doped electron at $K$ or $K'$ point
to form a trion leads to a Dirac type dispersion with a mass term,
coming from exchange coupling between exciton and electron,
$H_{\text{gap}}=\Delta \sigma_z s_z$, where $\sigma_i$ and $s_i$
represent valley
degrees of freedom of constituting exciton and electron,
respectively. The estimated value is $\Delta\sim 3 \,{\rm meV}$, so if
$ T\leq 17 \,{\rm K}$, our model with $\Omega \sim 6.9\times
10^{4}\,{\rm \AA^2}$
is applicable for the same reason as above.
Moreover, since trion is a charged particle, by applying magnetic field
$B = (\hbar/q)/\Omega\sim -970 \,{\rm mT}$, we expect the singular
behavior of $\gamma_1$ and $\gamma_2$ as we discussed above.
Here, $B$ is so small that we
can neglect the effect of Zeeman energy.
In both cases, for laser spot of $0.5 \,{\rm \mu m}$, the uncertainty in momentum
space is $\Delta k\sim 2 \times 10^{-4}\,{\rm \AA^{-1}}$, and well within
$\Delta/v$.
The time- and space-resolved spectra of light emission
can detect the diffusive dynamics of these particles.
The time-scale of the relaxation $\gamma^{-1}_{1,2}$
is typically pico second for electronic systems,
which is now within the range of
experimental access.

Besides above two, another candidate is the cold atom systems.
Recently, the topological band structure, i.e., Haldane model,
is realized in optical lattice \cite{Esslinger}.
It is also realized that the local defect is introduced as the initial
condition and trace the time-evolution of the system after it \cite{Fukuhara1,Fukuhara2}. 
In the case of cold atoms in optical lattice, the random force and dissipation 
is rather weak, and one needs to design the coupling of the atoms to 
the heat bath such as the electromagnetic field.
However, the time scale in this case is much longer, i.e., typically 
$\sim 10$msec \cite{Fukuhara2}, 
and the observation of the dynamics of a single particle 
is expected to be easier than the electronic systems.

Finally, we point out the difference between our work and the work in
the previous literature \cite{cobanera2016quantum,ozawa2014anomalous}.
In Ref. \onlinecite{ozawa2014anomalous}, the method of measuring the
momentum space Berry curvature in the lossy photonic lattice systems was
discussed. Although
the idea of measuring the Berry curvature through the optical excitation
and the resultant real-space distribution has some resemblance to our
proposal, there are important differences: They discussed the effect of
momentum space Berry curvature on a steady state
property (especially $\average{x}$)
of a lossy system with a continuous pumping at zero temperature, while
we discussed the effect of phase space Berry curvatures on the diffusive transient
dynamics (including the first and second moment in phase space)
after the irradiation of light at finite temperature.
In Ref. \onlinecite{cobanera2016quantum}, the diffusive dynamics of an
electron in a Landau level was discussed.
Although their treatment is
fully quantum mechanical and ours is semiclassical, our model is more
general when restricted to the semiclassical regime:
Since projecting onto a Landau
level corresponds to neglecting the kinetic term, their model in the
semiclassical, high temperature regime
corresponds to the special case of our model with $qB\neq 0$
and $m\to 0$ with $m\gamma$ fixed in Eq. (\ref{Langep}).

In summary, we find Berry curvatures modify
the relaxation time of the probability distribution in momentum space
and the diffusion coefficient.
In particular, the short time behavior contains useful information 
and hence the time-resolved experiments will provide useful information on 
Berry curvatures.
\begin{acknowledgments}
The authors thank M. Ezawa,
T. Fukuhara, S. Furukawa, T. Ideue, H. Ishizuka,
Y. Iwasa, M. Onga,
and M. Ueda for useful discussion.
This work was supported 
by the Elements Strategy Initiative
Center for Magnetic Materials (ESICMM) under the outsourcing
project of MEXT
(S.M.), and
Grants-in-Aid for Scientific Research (nos. 24224009 and 26103006)
from MEXT, Japan, and
ImPACT Program of Council for Science, Technology and
Innovation (Cabinet office, Government of Japan),
and JST CREST Grant Numbers
JPMJCR16F1, Japan (N.N.).
\end{acknowledgments}
\appendix
\section{Semiclassical equation in the presence of Berry curvatures}
 The physical meaning of each term in the semiclassical equation,
 Eqs. (\ref{Langer}) and (\ref{Langep})
is the followings.
$\epsilon(\bm{r},\bm{p})$ is the
energy of the particle and reflects the
potential energy and the dispersion relation of the band.
To understand the origin of the terms containing the Berry
curvature in the equation, it is important
to note that Berry connection is defined as the inner product of the
adjacent wave functions in some parameter space; here, the parameter
space is a phase space spanned by the position and momentum of a
particle. Since the particle is represented by the wave packet composed
of the neighboring wave functions,
the dynamics of the particle is affected by Berry connections.
The Berry connections appear in the
Lagrangian of the system, derived by the time dependent variational
principle \cite{Xiao2010}.
Except the last two terms in the right hand side of
Eq. (\ref{Langep}), Eqs. (\ref{Langer}) and (\ref{Langep})
are derived from the
effective Lagrangian of the system,
\begin{equation}
 L =  p_{i}\dot{r}^{i}+A_{i}(\bm{r},\bm{p})
  \dot{r}^{i}+a_{i}(\bm{r},\bm{p})\dot{p}^{i}-\epsilon(\bm{r},\bm{p}),
  \label{effectivelagrangian}
\end{equation}
where $A_{i}(\bm{r},\bm{p})$ and $a_{i}(\bm{r},\bm{p})$
are Berry connections of the wave function
in real space and momentum space, respectively. To see the role of each
term in the Lagrangian, we rewrite Eqs. (\ref{Langer}) and
(\ref{Langep}) as,
  \begin{equation}
\hat{G}^{-1}\left(
\begin{array}{c}
 \dot{\bm{r}}\\
 \dot{\bm{p}}\\
\end{array}
 \right)
 =-\left(
 \begin{array}{c}
  \nabla_{\bm{r}}\\
    \nabla_{\bm{p}}\\
   \end{array}
 \right)\epsilon(\bm{r},\bm{p})+\sqrt{2m\gamma k_B T}
 \left(
\begin{array}{c}
 \bm{\xi}(t)\\
 0\\
\end{array}
 \right), \label{Langmatrix}
\end{equation}
where
\begin{equation}
 \hat{G}^{-1}=\left(
\begin{array}{cc}
 m\gamma \hat{I}_d&0 \\
 0&0 \\
\end{array}
		      \right)
 +\left[\left(
\begin{array}{cc}
 0&\hat{I}_{d} \\
 -\hat{I}_{d}&0 \\
\end{array}
		      \right)
  -\left(
\begin{array}{cc}
 \hat{\Omega}_{rr}&\hat{\Omega}_{rp} \\
 \hat{\Omega}_{pr}&\hat{\Omega}_{pp} \\
\end{array}
 \right)\right];\label{Ginverse1}
\end{equation}
$\hat{I}_d$ is a $d \times d$ unit matrix;
the $d \times d$ matrices
$\hat{\Omega}_{rr},\hat{\Omega}_{rp},\hat{\Omega}_{pr}$ and $\hat{\Omega}_{pp}$
represent the
Berry curvatures and are defined as the field strengths in phase space:
\begin{align}
 &(\hat{\Omega}_{rr})_{ij}=\partial_{r_i}A_j-\partial_{r_j}A_i,\,
 (\hat{\Omega}_{rp})_{ij}=\partial_{r_i}a_j-\partial_{p_j}A_i,\\
 &(\hat{\Omega}_{pr})_{ij}=\partial_{p_i}A_j-\partial_{r_j}a_i,\,
 (\hat{\Omega}_{pp})_{ij}=\partial_{p_i}a_j-\partial_{p_j}a_i.
\end{align}
From the term in the square bracket in Eq. (\ref{Ginverse1}),
we can see that the first three terms in
Eq. (\ref{effectivelagrangian}) represent the
symplectic structure of the system. In particular,
the first term in the parenthesis on the right hand side of Eq.
(\ref{Langer}) is known
as a source of Hall effect, and
is called the anomalous velocity term \cite{nagaosa2010anomalous}.

\section{Langevin equation in the presence of Berry curvatures}
The Langevin equation of the particle with the energy
$\epsilon(\bm{r},\bm{p})$ in the presence of Berry curvatures is
\cite{ao2004potential,yin2006existence,olson2007nonequilibrium},
\begin{widetext}
\begin{equation}
 (\hat{G}^{-1})_{\alpha\beta}\dot{X}_{\beta}=-\nabla_{\alpha}\epsilon(\bm{X})+N_{\alpha\beta}\xi_{\beta}(t)
  \Leftrightarrow
  \dot{X}_{\alpha}=-G_{\alpha\beta}\nabla_{\beta}\epsilon(\bm{X})+
  (\hat{G}\hat{N})_{\alpha\beta}\xi_{\beta}(t),\label{Langevinsupp}
\end{equation}
\end{widetext}
where $\bm{X}=(\bm{r},\bm{p})$ and
\begin{align}
 \hat{G}^{-1}&=\hat{Q}
 +\left[\left(
\begin{array}{cc}
 0&\hat{I}_d \\
 -\hat{I}_d&0 \\
\end{array}
		      \right)
  -\left(
\begin{array}{cc}
 \hat{\Omega}_{rr}&\hat{\Omega}_{rp} \\
 \hat{\Omega}_{pr}&\hat{\Omega}_{pp} \\
\end{array}
 \right)\right].
 \label{Ginverse}
\end{align}
Here, $\hat{Q}$ is some $2d \times 2d$ symmetric matrix which represents the
effect of friction
and $\xi_{\alpha}(t)$ is the Gaussian fluctuation force:
\begin{equation}
 \average{\xi_{\alpha}(t)\xi_{\beta}(t')}=\delta(t-t')\delta_{\alpha\beta},
  \quad \average{\xi_{\alpha}(t)}=0.
\end{equation}
The subscript $\alpha,\beta=1,\dots,2d$ represent the coordinates of phase
space, $\hat{I}_d$ is
a $d \times d$ unit matrix.
This stochastic differential equation
does not necessarily describe the dynamics of a particle
coupled with a thermal bath; we need
to impose the condition which ensures the relaxation of the system
toward the equilibrium (Eq. (\ref{flucdiss})).
This condition can be derived from the
Fokker-Planck equation, which is equivalent to the Langevin equation
equipped with the interpretation of the noise term.
From now on, we assume that $\hat{G}\hat{N}$ does not depend on $\bm{X}$ to avoid
the subtlety of the interpretation of the noise term.
In general, given
some stochastic differential equation,
\begin{equation}
 \dot{x}_i(t)=g_i(\bm{x}(t))+h_{ij}\xi_j(t),
\end{equation}
 we can derive the time evolution of the probability distribution
 $P(\bm{r},t)=\average{\prod_i \delta(r_i-x_i(t))}$. First,
 \begin{widetext}
 \begin{align}
  x_i(t+\epsilon)&=x_i(t)+\int_{t}^{t+\epsilon}dt_1\, g_i(\bm{x}(t_1))
  +h_{ij}\int_{t}^{t+\epsilon}dt_1\,\xi_j(t_1). \label{discrete}
 \end{align}
 To evaluate this up to $\mathcal{O}(\epsilon)$, we note that for some
 arbitrary function $L(\bm{x}(t_1))$,
 \begin{align}
  L(\bm{x}(t_1))
  &=L(\bm{x}(t)+\bm{x}(t_1)-\bm{x}(t))\nonumber\\
  &=L(\bm{x}(t))+(x_k(t_1)-x_k(t))\nabla_k L(\bm{x}(t))+\dots \nonumber\\
  &=L(\bm{x}(t))+\int_t^{t_1}dt_2\, \dot{x}_k(t_2)\nabla_k
  L(\bm{x}(t))+\dots \nonumber\\
  &=L(\bm{x}(t))+\int_t^{t_1}dt_2\,
  \left[g_k(\bm{x}(t_2))+h_{kl}\xi_l(t_2)\right]\nabla_k
  L(\bm{x}(t))+\dots.
 \end{align}
 \end{widetext}
 To evaluate the order of the second term, we note that
 \begin{align}
  &\average{\int_t^{t+\epsilon}dt_1\,\xi_i(t_1)\int_t^{t+\epsilon}dt_1\,\xi_j(t_1)}\nonumber\\
  =&\int_t^{t+\epsilon}dt_1\int_t^{t+\epsilon}dt_2
   \average{\xi_i(t_1)\xi_j(t_2)}=\delta_{ij}\epsilon. \label{evaluatesupp}
 \end{align}
 So,
 \begin{equation}
  \int_t^{t+\epsilon}dt_1\,\xi_i(t_1)=\mathcal{O}(\epsilon^{\frac{1}{2}}). \label{xising}
 \end{equation}
 Then the right hand side of Eq. (\ref{discrete}) can be evaluated as
 \begin{align}
  x_i(t)+\epsilon g_i(\bm{x}(t))+\mathcal{O}(\epsilon^{\frac{3}{2}})+
  h_{ij}\int_t^{t+\epsilon}dt_1\, \xi_j(t_1).\label{evalsupp}
 \end{align}
 From Eq. (\ref{evalsupp}), we can calculate the first and second
 moments,
 \begin{widetext}
 \begin{align}
  &\lim_{\epsilon\to 0}
   \frac{1}{\epsilon}
   \average{x_i(t+\epsilon)-x_i(t)}=g_i(\bm{x}(t))\eqqcolon
  a_i(\bm{x}(t)),\\
  &\lim_{\epsilon\to 0}
  \frac{1}{\epsilon}
  \average{(x_i(t+\epsilon)-x_i(t))(x_j(t+\epsilon)-x_j(t))}=
   h_{ik}
   h_{jl}\delta_{kl}\eqqcolon a_{ij}(\bm{x}(t)).
 \end{align}
 And higher order moments are $\mathcal{O}(\epsilon^{\frac{3}{2}})$.
 From these moments, with the
 Chapman-Kolmogorov equation for this Markov process
 and its Kramers-Moyal expansion \cite{altland2010condensed},
 \begin{align}
  P(\bm{x},t+\epsilon)&=\int d\bm{x}'
   P(\bm{x},t+\epsilon|\bm{x}',t)P(\bm{x}',t)\\
   &=\int d\bm{x}'
  P((\bm{x}-\bm{x}')+\bm{x}',t+\epsilon|\bm{x}-\bm{x}',t)P(\bm{x}-\bm{x}',t)\nonumber \\
  &\eqqcolon \int d\bm{x}'
  F(\bm{x}-\bm{x}',\bm{x}';t+\epsilon,t)\nonumber\\
  &= \sum_{i_1,\dots,i_{D}}\frac{(-1)^{i_1+\dots +i_{D}}}{i_1!\dots i_{D}!}
  \left(\frac{\partial }{\partial
  x_1}\right)^{i_1}\dots \left(\frac{\partial }{\partial
  x_{D}}\right)^{i_{D}}\int d\bm{x}' x_1'^{i_1}\dots
  x_{D}'^{i_{D}}F(\bm{x}, \bm{x}';t+\epsilon,t)\\
  &=P(\bm{x},t)-\left(\frac{\partial }{\partial
  x_i}\right)(\epsilon a_i(\bm{x})P(\bm{x},t))
  +\frac{1}{2}\left(\frac{\partial^2 }{\partial
  x_i \partial x_j}\right)(\epsilon
  a_{ij}(\bm{x})P(\bm{x},t))+\mathcal{O}(\epsilon^{\frac{3}{2}}),
 \end{align}
 \end{widetext}
 where $D$ is the dimension of the system.
 So, if we take $\epsilon\to 0$, we obtain the Fokker-Planck equation:
 \begin{equation}
  \frac{\partial P(\bm{x},t)}{\partial t}=-
   \left(\frac{\partial}{\partial x_i}\right)(a_i(\bm{x})P)
   +\frac{1}{2}\left(
		\frac{\partial^2}{\partial x_i
		x_j}\right)(a_{ij}(\bm{x})P).\label{FPeqsupp}
 \end{equation}
 If we calculate the moments from Eq. (\ref{Langevinsupp}), we get
 \begin{align}
  a_{\alpha}(\bm{X})&=-G_{\alpha\beta}\nabla_{\beta}\epsilon,\\
  a_{\alpha\beta}(\bm{X})&=(\hat{G}\hat{N})_{\alpha\gamma}(\hat{G}\hat{N})_{\beta\delta}\delta_{\gamma\delta}=(\hat{G}\hat{N}\hat{N}^T\hat{G}^T)_{\alpha\beta}.
 \end{align}
 
 The system will eventually relax to the thermal
 equilibrium if the fluctuations and frictions are caused by a heat
 bath. From this physical assumption, we impose the condition that, the
 equilibrium distribution,
 \begin{equation}
  P_{\text{eq}}=\exp\left(-\frac{\epsilon(\bm{X})}{k_BT}\right),\label{thermal}
 \end{equation}
where $k_B$ is the Boltzmann constant and $T$ is the temperature, is the
stationary solution of Eq. (\ref{FPeqsupp}). We note that $\epsilon$ in
Eq. (\ref{thermal}) is the same as the one in Eq. (\ref{Langevinsupp}), since
the effect of Berry curvatures are the modification of the symplectic
structure of the system and the energy of the system
is not modified. As we assumed that Berry curvature terms are constant in
phase space, the modification of the density of states
\cite{xiao2005berry,xiao2005publisher}
is constant and can be ignored. From this condition,
\begin{widetext}
\begin{align}
 &0=\left(\frac{\partial}{\partial X_{\alpha}}\right)
 (G_{\alpha\beta}\nabla_{\beta}\epsilon P_{\text{eq}})
   +\frac{1}{2}\left(
		\frac{\partial^2}{\partial X_{\alpha}
 X_{\beta}}\right)((\hat{G}\hat{N}\hat{N}^T\hat{G})_{\alpha\beta}P_{\text{eq}})\\
 \Leftrightarrow &0=(\nabla_{\alpha}\nabla_{\beta}\epsilon)
 \left(G_{\alpha\beta}-\frac{1}{2k_BT}(\hat{G}\hat{N}\hat{N}^T\hat{G}^T)_{\alpha\beta}\right)\nonumber\\
 &\qquad+(\nabla_{\alpha}\epsilon)(\nabla_{\beta}\epsilon)\left(-\frac{1}{k_BT}G_{\alpha\beta}+\frac{1}{2(k_BT)^2}(\hat{G}\hat{N}\hat{N}^T\hat{G}^T)_{\alpha\beta}\right).
\end{align}
\end{widetext}
As a result, we obtain the condition
\begin{equation}
\frac{1}{2}\left(\hat{G}^{-1}+(\hat{G}^{-1})^T\right)=\frac{1}{2k_BT}\hat{N}\hat{N}^T.
  \label{flucdiss}
\end{equation}
This condition relates friction terms to fluctuation terms, and
is called the fluctuation-dissipation relationship.

Up to now, as far as the condition Eq. (\ref{flucdiss}) is
satisfied, we can
choose arbitrary form for $\hat{Q}$ and $\hat{N}$. Here we consider the microscopic
derivation of the Langevin equation (\ref{Langevinsupp}) by coupling
the system with a bath to decide the form of $\hat{Q}$ and $\hat{N}$ in that situation.
\section{Derivation of Eq. (\ref{Langevinsupp}) from Feynman and Vernon's influential functional}
To derive the form of friction and fluctuation terms in Eq. (\ref{Langevinsupp}),
we consider the Caldeira-Leggett model \cite{caldeira1983path,kamenev2011field}
in the presence of Berry curvatures. The argument here closely follows
the one in Ref. \onlinecite{kamenev2011field}.
 We set $\hbar=1$ and $k_B=1$ in
this section.
The action of the system is,
\begin{align}
 S_{\text{sys}}=\int_{C}d\tau \,\left(p_{i}\dot{r}_{i}+A_{i}(\bm{r},\bm{p})
  \dot{r}_{i}+a_{i}(\bm{r},\bm{p})\dot{p}_{i}-\frac{p_i^2}{2m}\right),
\end{align}
where $C$ is the closed time contour and
$C=C_{+}\cup C_{-}=\{t_i+i0,t_f+i0\}\cup\{t_f-i0,t_i-i0\}$.
Here we
consider two dimensional system and the form of Berry curvatures are
\begin{align}
&\hat{\Omega}_{pp}=\left(
\begin{array}{cc}
 0&\Omega \\
 -\Omega&0 \\
\end{array}
			 \right),\quad
  \hat{\Omega}_{rr}=\left(
\begin{array}{cc}
 0&qB \\
 -qB&0 \\
\end{array}
 \right),\nonumber\\
 &\hat{\Omega}_{rp}=\left(
\begin{array}{cc}
 C&A \\
 -A&C \\
\end{array}
			 \right),\quad
 \hat{\Omega}_{pr}=\left(
\begin{array}{cc}
 -C&A \\
 -A&-C \\
\end{array}
			 \right).\label{Berrycurv1}
\end{align}
Then,
\begin{equation}
 a_i=\frac{\Omega}{2}\epsilon_{ji}p_j+A\epsilon_{ji}r_j,\quad
  A_i=\frac{qB}{2}\epsilon_{ji}r_j-Cp_i,
\end{equation}
where $\epsilon_{ji}$ is the antisymmetric tensor.
We define $r_i(t+i0
)\eqqcolon r_i^{+}(t)$, $r_i(t-i0)\eqqcolon r_i^{-}(t)$, and
$r_i^{\text{cl}(\text{q})}(t)\eqqcolon \frac{1}{2}(r_i^{+}(t)\pm
r_i^{-}(t))$.
We use the same
definition also for all the fields in the Keldysh space. Then we get
\begin{widetext}
\begin{align}
 S_{\text{sys}}=2\int_{t_i}^{t_f}
 d\tau \,&\Biggl(p^{\text{q}}_{i}\dot{r}_{i}^{\text{cl}}+
 p^{\text{cl}}_{i}\dot{r}_{i}^{\text{q}}+
	     qB\epsilon_{ji}r^{\text{q}}_j
	     \dot{r}_{i}^{\text{cl}}
	     +\Omega\epsilon_{ji}p_j^{\text{q}}
 \dot{p}_{i}^{\text{cl}}\nonumber\\
&\quad+A\epsilon_{ji}r_j^{\text{q}}
\dot{p}_{i}^{\text{cl}}+A\epsilon_{ji}r_j^{\text{cl}}
\dot{p}_{i}^{\text{q}}-Cp^{\text{q}}_i\dot{r}_i^{\text{cl}}-
Cp^{\text{cl}}_i\dot{r}_i^{\text{q}}
	     -\frac{p_i^{\text{cl}}
	     p_i^{\text{q}}}{m}\Biggr).
\end{align}
Here, we couple the system with a bath which is a collection of
oscillators labeled by $s$ \cite{kamenev2011field}:
\begin{align}
 &S_{\text{bath}}=\frac{1}{2}\sum_{s,i}\int_{-\infty}^{+\infty}
 dt \, \vec{\phi}_{s,i}^{T}(t)\hat{D}_s^{-1}(t)\vec{\phi}_{s,i}(t),\\
 &S_{\text{int}}=\sum_{s,i}
 g_s\int_{-\infty}^{+\infty}dt\, \left(r_i^{+}\phi_{s,i}^{+}
 -r_i^{-}\phi_{s,i}^{-}\right)
 =\sum_{s,i}
 2 g_s\int_{-\infty}^{+\infty}dt\, \vec{r}_i^T(t) \hat{\sigma}_x
 \vec{\phi}_{s,i}(t),
\end{align}
\end{widetext}
where $\hat{\sigma}_x$ is the $x$ component of the Pauli matrix
in Keldysh space;
the vector represents
\begin{equation}
 \vec{r}_i^{T}=(r_i^{\text{cl}},r_i^{\text{q}}),\quad
  \vec{\phi}_{s,i}^T=(\phi_{s,i}^{\text{cl}},\phi_{s,i}^{\text{q}});
\end{equation}
$\hat{D}^{-1}_s$ is a $2 \times 2$ matrix in Keldysh space:
\begin{align}
 &\hat{D}^{-1}_s(t)=\left(
\begin{array}{cc}
 0&{[D_s^{-1}}]^{\text{A}}(t) \\
 {[D_s^{-1}}]^{\text{R}}(t) &{[D_s^{-1}}]^{\text{K}}(t)\\
\end{array}
		       \right),\nonumber\\
 &\hat{D}_s(t)=\left(
\begin{array}{cc}
 D_s^{\text{K}}(t)&D_s^{\text{R}}(t) \\
 D_s^{\text{A}}(t) &0\\
\end{array}
		       \right).
\end{align}
And from the dispersion relationship of the harmonic
oscillator and the fluctuation-dissipation relationship for the heat
bath, in the Fourier transformed basis,
\begin{align}
 D_s^{\text{R}(\text{A})}(\epsilon)&=\frac{1}{2}\frac{1}{(\epsilon\pm
  i0)^2-\omega_s^2},\nonumber\\
  D_s^{\text{K}}(\epsilon)&=\coth
  \frac{\epsilon}{2T}
  \left[D^{\text{R}}_s(\epsilon)-D^{\text{A}}_s(\epsilon)\right]\nonumber\\
  &\cong \frac{2T}{\epsilon}
  \left[D^{\text{R}}_s(\epsilon)-D^{\text{A}}_s(\epsilon)\right],
\end{align}
where in the last equation, we assume the temperature is high compared
to the characteristic frequency of the oscillator
(semiclassical approximation); $\omega_s$ is the frequency of the
oscillator $s$.

If we trace them out, there remains the terms which
represent the interaction between
forward and backward contours of the system. These terms are called the
influence functional \cite{feynman1963theory}.
Since the argument is exactly the same as the model in the
absence of Berry curvatures
\cite{kamenev2011field},
we just show the results.
If we assume the Ohmic bath:
\begin{equation}
 J(\omega)\coloneqq
  \pi\sum_{s}\frac{g_s^2}{\omega_s}\delta(\omega-\omega_s)
  =2m\gamma\omega,
\end{equation}
the contribution of the bath to the
effective action for the system coordinate is,
\begin{widetext}
\begin{align}
 S_{\text{int}}&=\frac{1}{2}\int\int_{-\infty}^{+\infty}dt\,dt'\,
 \sum_i\vec{r}_i^T(t) \left[-\sum_s (2g_s)^2\hat{\sigma}_x \hat{D}_s(t-t')
 \hat{\sigma}_x\right]
 \vec{r}_i(t')\nonumber\\
 &\eqqcolon
 \frac{1}{2}\int\int_{-\infty}^{+\infty}dt\,dt'\,
 \sum_i\vec{r}_i^T(t) \hat{\mathfrak{D}}^{-1}(t-t')
 \vec{r}_i(t').
\end{align}
\end{widetext}
Since
\begin{align}
 [\mathfrak{D}^{-1}(\epsilon)]^{\text{R}(\text{A})}
  &=-\frac{1}{2}\sum_{s}\frac{4g_s^2}{(\epsilon\pm i0)-\omega_s^2}\nonumber\\
  &=\int_0^{+\infty} \frac{d\omega}{2\pi}
  \frac{4\omega J(\omega)}{\omega^2 -(\epsilon \pm i0)^2}
 =R\pm 2i m\gamma\epsilon,\nonumber\\
  [\mathfrak{D}^{-1}(\epsilon)]^{\text{K}}
  &\cong
 \left([\mathfrak{D}^{-1}(\epsilon)]^{\text{R}}-
 [\mathfrak{D}^{-1}(\epsilon)]^{\text{A}}\right)
  \frac{2T}{\epsilon}=8im\gamma T,
\end{align}
where the constant real part of
$[\mathfrak{D}^{-1}(\epsilon)]^{\text{R}(\text{A})}$,
$R$ renormalizes the potential of the particle,
and we will ignore this term. Then, after Fourier transforming back to
the time representation,
\begin{equation}
 S_{\text{int}}=-2m\gamma\int dt r_i^{\text{q}}\dot{r}_i^{\text{cl}}+
4im\gamma T \int dt (r_i^{\text{q}})^2.
\end{equation}
The second term can be rewritten as
\begin{equation}
 e^{-4m\gamma T\int dt (r_i^{\text{q}})^2}=\int D\left[\xi_i(t)\right]e^{
  -\int dt\,\left[\frac{\xi_i(t)^2}{4m\gamma
	     T}-2i\xi_i(t)r_i^{\text{q}}(t)\right]}. 
\end{equation}
As a result, after performing $r_i^{\text{q}}$ and $p_i^{\text{q}}$
integration, we get the expression for the expectation value of
the observable $\Omega(\bm{r^{\text{cl}}},\bm{p^{\text{cl}}})$
\begin{widetext}
 \begin{align}
  &\average{\Omega(\bm{r^{\text{cl}}}(t),\bm{p^{\text{cl}}}(t))}
 =\int D\left[\xi_i(t)\right] e^{-\int dt\,\frac{1}{4m\gamma k_BT}\xi_i(t)^2}
  \int
  D\left[r_i^{\text{cl}}(t)p_i^{\text{cl}}(t)\right]
  \Omega(\bm{r^{\text{cl}}}(t),\bm{p^{\text{cl}}}(t))\nonumber\\
  \times&\prod_i\delta(\dot{r}_i^{\text{cl}}-\frac{p_i^{\text{cl}}}{m}+\Omega
  \epsilon_{ij}\dot{r}_j^{\text{cl}}+A\epsilon_{ij}\dot{r}_j^{\text{cl}}
  -C\dot{r}^{\text{cl}}_i)
  \delta(\dot{p}_i^{\text{cl}}-qB\epsilon_{ij}\dot{r}_j^{\text{cl}}
  -A\epsilon_{ij}\dot{p}_j^{\text{cl}}-C\dot{p}^{\text{cl}}_i+m\gamma
 \dot{r}_i^{\text{cl}}-\xi_i),
 \end{align}
\end{widetext}
 where we set $T\to k_B T$.
This expression represents the Langevin equation (\ref{Langevinsupp})
with
\begin{equation}
  \hat{N}=\left(
\begin{array}{cc}
 \sqrt{2m\gamma k_BT}\hat{I}_2&0 \\
 0&0 \\
\end{array}
	       \right),\quad
 \hat{Q}=\left(
\begin{array}{cc}
 m\gamma \hat{I}_2&0 \\
 0&0 \\
\end{array}
	       \right),\label{Asupp}
\end{equation}
where $\bm{X}=(\bm{r},\bm{p})$.
Then, Eq. (\ref{Langevinsupp}) is nothing but Eqs. (\ref{Langer}) and (\ref{Langep})

Here we note that, the friction term on the right hand side of
Eq. (\ref{Langep}) is $-m\gamma \dot{r}_i$,
not $-\gamma p_i$. The reason is that, the friction term $-\gamma p_i$
and
the fluctuation
term $\sqrt{2m\gamma k_BT}\xi_i(t)$
do not satisfy Eq. (\ref{flucdiss}).
Also, the microscopic derivation above leads
to the friction term $-m\gamma \dot{r}_i$ and the fluctuation term
$\sqrt{2m\gamma k_BT}\xi_i(t)$, which satisfy equation
(\ref{flucdiss}). Therefore, as far as this microscopic model is valid
for the description of the dynamics, Eqs. (\ref{Langer}) and
(\ref{Langep}) must be used.

From now on, we will use Eq. (\ref{Asupp}). Then, from Eq. (\ref{flucdiss}),
Eq. (\ref{FPeqsupp}) can be rewritten as
\begin{align}
  \frac{\partial P(\bm{X},t)}{\partial
 t}&=G_{\alpha\beta}\nabla_{\alpha}((\nabla_{\beta}\epsilon)P)\nonumber\\
 &\qquad+\frac{k_BT}{2}(\hat{G}+\hat{G}^T)_{\alpha\beta}\nabla_{\alpha}\nabla_{\beta}P.
 \label{FPhon}
\end{align}
 \section{Exact results}
 Given any linear multivariate Fokker-Planck equation,
 \begin{equation}
  \frac{\partial P(\bm{X},t)}{\partial t}
   =-A_{\alpha\beta}\frac{\partial}{\partial
   X_{\alpha}}(X_{\beta}P)+\frac{1}{2}B_{\alpha\beta}\frac{\partial^2 P}{\partial X_{\alpha}\partial X_{\beta}},\label{linearFPsupp}
 \end{equation}
 where $\hat{A}$ and $\hat{B}$ are the constant matrices, we can exactly solve it
 with
 the initial condition  \cite{van1992stochastic}
 \begin{equation}
  P(\bm{X},0)=\prod_{i=1}^{2d}\delta(X_i-X_{i0}).
 \end{equation}
 If we multiply Eq. (\ref{linearFPsupp}) with $X_\gamma$ and integrate over $\bm{X}$,
 we get
 \begin{equation}
  \frac{\partial}{\partial t}\average{X_\gamma}=A_{\gamma \beta}
   \average{X_{\beta}},
 \end{equation}
 then
 \begin{equation}
  \average{X_\gamma}(t)=(\exp(t\hat{A}))_{\gamma \beta}X_{\beta0}.
   \label{averagesupp}
 \end{equation}
 If we multiply Eq. (\ref{linearFPsupp}) with $X_\gamma X_\delta$ and integrate over $\bm{X}$,
 we get
 \begin{equation}
  \frac{\partial}{\partial t}\average{X_\gamma X_\delta}=A_{\gamma
   \alpha}\average{X_\alpha X_\delta}
   +A_{\delta\beta}\average{X_\gamma X_\beta}+B_{\gamma \delta}.
 \end{equation}
 If we introduce
 \begin{equation}
  \average{\average{X_\gamma(t)X_\delta(t)}}=\average{X_\gamma X_\delta}(t)-\average{X_{\gamma}}(t)\average{X_\delta}(t)\eqqcolon
   \Theta_{\gamma \delta}(t),
 \end{equation}
 \begin{equation}
  \hat{\Theta}^*(t)\coloneqq e^{-t\hat{A}}\hat{\Theta}(t)e^{-t\hat{A}^T},
 \end{equation}
 then $\Theta^*_{\gamma \delta}(0)=0$ and
 \begin{equation}
  \frac{\partial}{\partial t}\hat{\Theta}^*=e^{-t\hat{A}}\hat{B}e^{-t\hat{A}^T}.
 \end{equation}
 As a result, we get
 \begin{align}
  &\hat{\Theta}^*(t)=\int_0^t dt'\,
  e^{-t'\hat{A}}\hat{B}e^{-t'\hat{A}^T}\nonumber\\
  \Leftrightarrow &\hat{\Theta}(t)=\int_0^t dt'\,
  e^{(t-t')\hat{A}}\hat{B}e^{(t-t')\hat{A}^T}=
  \int_0^t dt'\, e^{t'\hat{A}}\hat{B}e^{t'\hat{A}^T}.\label{covariantsupp}
 \end{align}
 Eqs. (\ref{averagesupp}) and (\ref{covariantsupp}), are enough
 to determine the whole dynamics since the process is Gaussian. The
 solution is,
 \begin{align}
  &P(\bm{X},t)=(2\pi)^{-d}(\text{det}\hat{\Theta})^{-\frac{1}{2}}\nonumber\\
  &\times\exp\left[
							       -\frac{1}{2}(\bm{X}^T-\average{\bm{X}^T}(t))
	\hat{\Theta}^{-1}(t)(\bm{X}-\average{\bm{X}}(t))\right].
 \end{align}

 If we set $\epsilon(\bm{p})=\bm{p}^2/(2m)$, the Fokker-Planck equation
 with Berry curvatures, Eq. (\ref{FPhon}), are linear multivariate
 and
 \begin{align}
  \hat{A}&=\frac{1}{m}\left(
\begin{array}{cc}
 0&-\hat{G}_{\bm{r}\bm{p}} \\
 0&-\hat{G}_{\bm{p}\bm{p}} \\
\end{array}
  \right),\label{Asupple}\\
  B_{\alpha\beta}&=k_BT(G_{\alpha\beta}+G_{\beta\alpha}),\label{Bsupple}
 \end{align}
 where
 \begin{align}
  \hat{G}&\eqqcolon \left(
\begin{array}{cc}
 \hat{G}_{\bm{r}\bm{r}}&\hat{G}_{\bm{r}\bm{p}} \\
 \hat{G}_{\bm{p}\bm{r}}&\hat{G}_{\bm{p}\bm{p}} \\
\end{array}
			      \right),
 \end{align}
 \begin{widetext}
 \begin{align}
  \hat{G}_{\bm{r}\bm{r}}&=M\left[\frac{m\gamma\Omega^2}{D^2+A^2-qB\Omega} \hat{I}_2
  -\Omega i\hat{\sigma}_y\right],\\
  \hat{G}_{\bm{r}\bm{p}}&=M\left[\left(-D-
  \frac{m\gamma\Omega A}{D^2+A^2-qB\Omega}\right)\hat{I}_2+\left(A-\frac{m\gamma\Omega D}
  {D^2+A^2-qB\Omega} \right)i\hat{\sigma}_y\right], \\
  \hat{G}_{\bm{p}\bm{r}}&=M\left[\left(D-\frac{m\gamma\Omega
  A}{D^2+A^2-qB\Omega}\right)\hat{I}_2+
  \left(A+\frac{m\gamma\Omega D}{D^2+A^2-qB\Omega}\right)i\hat{\sigma}_y\right], \\
  \hat{G}_{\bm{p}\bm{p}}&=M\left[\frac{m\gamma(A^2+D^2)}{D^2+A^2-qB\Omega}
  \hat{I}_2+\left(-qB+\frac{m^2\gamma^2\Omega}{D^2+A^2-qB\Omega}\right)i\hat{\sigma}_y\right],
 \end{align}
 $D\coloneqq 1-C$ and
\begin{equation}
 M=\frac{D^2+A^2-qB\Omega}{(D^2+A^2-qB\Omega)^2+m^2\gamma^2\Omega^2}.
\end{equation}
 So we just need to calculate Eqs. (\ref{averagesupp}) and (\ref{covariantsupp})
 with matrices Eqs. (\ref{Asupple}) and (\ref{Bsupple}). 
 If we define
 \begin{equation}
  \gamma_1=\frac{(D^2+A^2)\gamma}{(D^2+A^2-qB\Omega)^2+m^2\gamma^2\Omega^2},
   \quad
   \gamma_2=\frac{-qB(D^2+A^2)+(q^2B^2+m^2\gamma^2)\Omega}{m[(D^2+A^2-qB\Omega)^2+m^2\gamma^2\Omega^2]},
 \end{equation}
 and
  \begin{align}
  &g_1(t)=\frac{1}{q^2B^2+m^2\gamma^2}[AqB+m\gamma D\nonumber\\
   &\qquad\qquad -(AqB+m\gamma
   D)e^{-\gamma_1 t}\cos(\gamma_2 t)+(m\gamma A-qBD)e^{-\gamma_1 t}
   \sin(\gamma_2 t)],\\
   &g_2(t)=\frac{1}{q^2B^2+m^2\gamma^2}[qBD-m\gamma A\nonumber \\
   &\qquad\qquad -(qBD-m\gamma A
   )e^{-\gamma_1 t}\cos(\gamma_2 t)+(m\gamma D+AqB)e^{-\gamma_1 t}
   \sin(\gamma_2 t)],\\
  &f_1(t)=m\gamma D-AqB+e^{-2\gamma_1 t}[m\gamma D+AqB-2m\gamma e^{\gamma_1 t}
  (D\cos(\gamma_2 t)+A\sin(\gamma_2 t))],\\
  &f_2(t)=m\gamma A+qBD-e^{-2\gamma_1 t}[qBD-m\gamma A+2m\gamma e^{\gamma_1 t}
  (A\cos(\gamma_2 t)-D\sin(\gamma_2 t))],
  \end{align}
 then Eqs. (\ref{averagesupp}) and (\ref{covariantsupp}) are,
  \begin{align}
   &\left(
\begin{array}{c}
 \average{p_x(t)}\\
 \average{p_y(t)}\\
\end{array}
	\right)
 =e^{-\gamma_1 t}
\left(
 \begin{array}{cc}
\cos(\gamma_2 t)&-\sin(\gamma_2 t)\\
 \sin(\gamma_2 t)&\cos(\gamma_2 t) \\
 \end{array}
 \right)
 \left(
\begin{array}{c}
 p_{x0}\\
 p_{y0}\\
\end{array}
   \right),\\
   &\left(
\begin{array}{c}
 \average{r_x(t)}\\
 \average{r_y(t)}\\
\end{array}
	\right)=
\left(
 \begin{array}{cc}
g_1(t)&g_2(t)\\
 -g_2(t)&g_1(t) \\
 \end{array}
 \right)
 \left(
\begin{array}{c}
 p_{x0}\\
 p_{y0}\\
\end{array}
   \right)+
  \left(
\begin{array}{c}
 r_{x0}\\
 r_{y0}\\
\end{array}
   \right) ,\\
  &\average{\average{r_i(t)r_j(t)}}=\biggl[\frac{2m\gamma
  k_BT}{q^2B^2+m^2\gamma^2}t+
  \frac{mk_BT(A^2+D^2)}{(q^2B^2+m^2\gamma^2)^2}(B^2-3m^2\gamma^2)\nonumber\\
  &\quad+\frac{4m^2\gamma k_BT(A^2+D^2)}{(q^2B^2+m^2\gamma^2)^2}
  e^{-\gamma_1 t}(m\gamma\cos(\gamma_2t)+qB\sin(\gamma_2 t))-
  \frac{mk_BT(A^2+D^2)}{q^2B^2+m^2\gamma^2}e^{-2\gamma_1 t}\biggr]\delta_{ij},\\
  &\average{\average{r_i(t)p_j(t)}}=\frac{mk_BT}{q^2B^2+m^2\gamma^2}
  (f_1(t)\delta_{ij}-f_2(t)(i\hat{\sigma}_y)_{ij}),\\
  &\average{\average{p_i(t)p_j(t)}}=mk_B T(1-e^{-2\gamma_1 t})\delta_{ij}.
  \end{align}
 In the main text, we
 put $A=0$. We note that,
 \begin{equation}
  D^2+A^2-qB\Omega=1-2C+C^2+A^2-qB\Omega=1-(\hat{\Omega}_{rp})_{ii}
   -\epsilon
   _{\alpha \beta \gamma \delta}(\hat{\Omega}_{X X})_{\alpha\beta}
   (\hat{\Omega}_{XX})_{\gamma\delta}/8,
 \end{equation}
 where $\epsilon_{\alpha \beta \gamma \delta}$ is the completely antisymmetric tensor,
 is nothing but the modified density of state of the system
 \cite{xiao2005berry,xiao2005publisher,Hayata2016kinetic}.
 \end{widetext}


%

\end{document}